\documentclass[%
 notitlepage,
 twocolumn,
superscriptaddress,
 amsmath,
 amssymb,
 aps,
 pra,
]{revtex4-1}
\usepackage{amsfonts}
\usepackage{graphicx}
\usepackage{physics}
\usepackage{bm}
\usepackage{soul}
\usepackage{color}
\usepackage{xcolor}
\usepackage{dsfont}
\usepackage{hyperref}
\hypersetup{
citecolor=blue,
 colorlinks=true,
 urlcolor=magenta,
}

\newcommand{\ICFO}{ICFO - Institut de Ciencies Fotoniques, The Barcelona Institute of Science and Technology, Av. Carl Friedrich Gauss 3, 08860 Castelldefels (Barcelona), Spain}

\newcommand{\IOPPAS}{Institute of Physics PAS, Aleja Lotnikow 32/46, 02-668 Warszawa, Poland}

\newcommand{\VILNIUS}{Institute of Theoretical Physics and Astronomy,
Vilnius University, Saul\.etekio 3, LT-10257, Vilnius, Lithuania}

\begin{document}

\title{
One- and two-axis squeezing via laser coupling in an atomic Fermi-Hubbard model
}

\author{T. Hern\'andez Yanes}
\affiliation{\IOPPAS}

\author{M. P\l{}odzie\'n}
\affiliation{\ICFO}

\author{M. Mackoit Sinkevi\v{c}ien\.e}
\affiliation{\VILNIUS}

\author{G. \v{Z}labys}
\affiliation{\VILNIUS}

\author{G. Juzeli\=unas}
\affiliation{\VILNIUS}

\author{E. Witkowska}
\affiliation{\IOPPAS}

\date{\today}

\begin{abstract}
Generation, storage, and utilization of correlated many-body quantum states are crucial objectives of future quantum technologies and metrology. Such states can be generated by the spin squeezing protocols,~i.e., one-axis twisting and two-axis counter-twisting. In this work, we show activation of these two squeezing mechanisms in a system composed of ultra-cold atomic fermions in the Mott insulating phase by a position-dependent laser coupling of atomic internal states. Realization of both the squeezing protocols is feasible in the current state-of-the-art experiments.
\end{abstract}

\maketitle

\emph{Introduction.-} 
The fundamental interest of emerging quantum technologies lies in many-body entangled and Bell correlated states, their production, and storage~\cite{Acin_2018,Kinos2021,Zwiller2022,Fraxanet2022}. Spin squeezing protocols, i.e. one-axis twisting (OAT) and two-axis counter-twisting (TACT), represents such a resource \cite{PhysRevA.47.5138, PhysRevA.50.67} particularly useful for high-precision measurements, allowing to overcome the shot-noise limit~\cite{RevModPhys.90.035005, PhysRevLett.105.053601, Chabuda2020}, and study many-body entanglement ~\cite{Fadel2018, PhysRevLett.122.173601,Hosten2016, Pedrozo2020, Bao2020} and Bell correlations~\cite{Tura1256,schmied2016bell,Aloy2019, Baccari2019, Tura2019, PRXQuantum.2.030329,Panfil2020,Plodzien2022}.
It applies to systems composed of $N$ particles in two quantum states described by a spin of quantum number $S=N/2$, when the final measurement is performed by spectroscopic experiments. The uncertainty of such a measurement is $\xi /\sqrt{N}$, where $\xi$ is the spin squeezing parameter~\cite{PhysRevA.50.67}, and wherein the shot-noise limit is $1/\sqrt{N}$. 
Simulation of OAT by means of unitary evolution requires a non-linear term in the system Hamiltonian. Such a term can be cast in the form $\hat{S}_z^2$ with $z$ being the twisting axis. TACT is a natural extension of the OAT model where the clockwise and counter-clockwise twisting take place around two orthogonal axis, such as $z$ and $x$, and can be described by the term $\hat{S}_z^2 - \hat{S}_x^2$. 
The lowest value of the squeezing parameter is $\xi_{\rm best} \propto N^{-1/3}$ for OAT and $\xi_{\rm best} \propto N^{-1/2}$ for TACT.

From an experimental point of view an important aspect in simulation of OAT and TACT is a well controlled quantum many-body system.
Several methods were proposed with  ultra-cold atoms, through the quantum non-demolition measurements~\cite{PhysRevLett.85.1594}, transfer of squeezing from light to atomic ensembles~\cite{PhysRevLett.83.1319} or 
utilizing atom-atom interactions~\cite{Schleier2010}. 
Proof-of-principle experiments were already performed to demonstrate spin squeezing via OAT with Bose-Einstein condensates utilizing atom-atom collisions
~\cite{Treutlein2010,Oberthaler2010,Chapman2012,PhysRevLett.125.033401} and atom-light interactions in cavity setups~\cite{PhysRevLett.104.073602,PhysRevLett.105.080403}. On the contrary, TACT was not demonstrated experimentally yet, despite several methods having been proposed~\cite{Molmer1999_PRL,PhysRevA.91.043642, PhysRevLett.107.013601, PhysRevA.96.013823, PhysRevResearch.2.033504, Borregaard_2017, PhysRevA.96.013823, Huang2021}. Currently, intensive research is carried out in lattice systems in the context of atomic lattice clocks~\cite{Sorensen1999PRL,Kajtoch-sc-2018,PhysRevResearch.1.033075, PhysRevA.102.013328, Vladan2020, PhysRevResearch.3.013178, PhysRevA.105.022625}. 
The drawbacks of spin squeezing generation with bosonic atoms are collision decoherence processes and losses~\cite{RevModPhys.90.035005}. An alternative is offered by spinful fermions in the Mott insulating phase where each lattice site is occupied by a single atom, and hence, the collision decoherence processes are reduced~\cite{Kolkowitz2017}. 

\begin{figure}
    \centering
    \includegraphics[width=\linewidth]{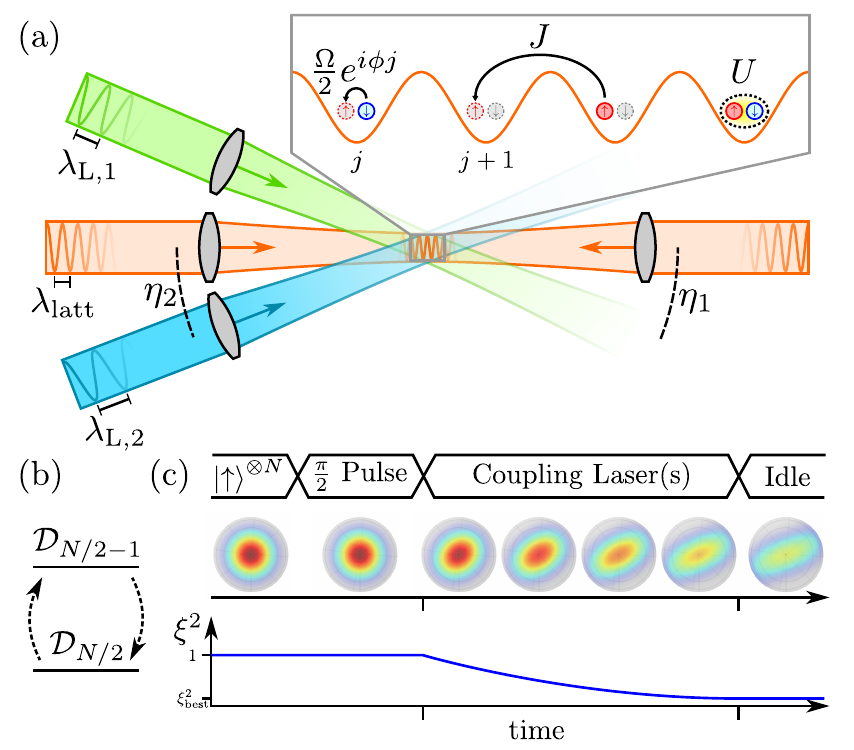}
     \caption{
     (a) Fermi-Hubbard model (FHM) for atoms in optical lattices with  nearest-neighbor tunneling rate $J$, on-site interaction $U$ and additional coupling between atomic internal degrees of freedom with position-dependent strength $\Omega e^{i \phi j}$ realized with one or two off-resonant laser beams. 
     (b) The coupling causes the transition between the ${\cal D}_{N/2}$ and ${\cal D}_{N/2-1}$ spin manifolds (see main text). In the weakly coupling regime, the projection of the Hamiltonian onto the Dicke manifold ${\cal D}_{N/2}$ leads to the OAT (single beam) and TACT (two beams) models. 
     (c) The Ramsey-type spectroscopy for the spin-squeezing generation: $(i)$ preparation of the initial spin coherent state in the ${\cal D}_{N/2}$ manifold, $(ii)$ unitary evolution with the FHM and non-zero coupling reduces the value of $\xi^2$, $(iii)$ storing the spin squeezed state in the Mott phase for zero atom-light coupling.
    }
\label{fig:fig1}
\end{figure}

In this letter, we study theoretically a scheme for the dynamical generation of both OAT and TACT by a position-dependent laser coupling of two atomic internal states, denoted by $\ket{\uparrow}$ and $\ket{\downarrow}$, of $N$ atomic fermions in a lattice, as illustrated in Fig.~\ref{fig:fig1}~(a). We describe the system by the Fermi-Hubbard model (FHM).
Additionally, there is a coupling between two internal states of atoms, \textit{i.e.} Raman coupling or direct optical transitions, which effectively acts as the spin-orbit coupling in the momentum representation~\cite{2009Spielman,2011Spielman,2011Fu,2012Zhang,2012Wang,2012Cheuk,2012Windpassinger,2012Garcia,2016JYe,Gediminas2011}. Generation of OAT due to such coupling was proposed for trapped ions~\cite{PhysRevLett.82.1835} and ultra-cold fermions following additional site-dependent spin rotations~\cite{PhysRevResearch.1.033075,PhysRevResearch.3.013178}.
Here, we show how not only the OAT but also TACT can be simulated directly from the FHM without any additional manipulation of individual spins.

We study the Ramsey-type spectroscopy scheme in which the coupling between atomic internal degrees of freedom is turned on during the interrogation time, as sketched in Fig.~\ref{fig:fig1}~(c). 
The generation of spin squeezing starts after the preparation of an initial spin coherent state which subsequently undergoes a unitary evolution. The spin squeezing parameter $\xi^2 = N (\Delta S_{\perp})^2_{\mathrm{min}}/\langle S \rangle^2$~\cite{PhysRevA.50.67} is then measured, where $\langle S \rangle$ is the length of the mean collective spin and $(\Delta S_{\perp})^2_{\mathrm{min}}$ is the minimal variance of the collective spin orthogonally to its direction. We show
analytically that the FHM in the Mott regime, supplied with the weak atom-light coupling, simulate both twisting models. Specifically, applying a single laser beam, the atom-light coupling induces OAT with a tunable axis of squeezing. This paves the way for the simulation of the famous TACT model~\cite{PhysRevA.47.5138,PhysRevA.92.013623} when two laser couplings are used during the interrogation time. To illustrate and demonstrate the validity of our analytical finding we perform full many-body calculations
\footnote{Our numerical calculations are based on the exact diagonalization method in the full Fock state basis for total number of atoms $N\leq 10$, without fixed magnetization. For larger number of atoms 
we employed a standard density matrix renormalization group (DMRG) technique for calculating initial spin coherent state \cite{PhysRevLett.69.2863,PhysRevB.48.10345,RevModPhys.77.259,SCHOLLWOCK201196,ORUS2014117}. Time evolution was prepared within algorithm for time evolution where one-site time-dependent variational principle (TDVP) scheme \cite{Kramer_2008, PhysRevLett.107.070601, PhysRevLett.109.267203,PhysRevB.94.165116} is combined with a global basis expansion \cite{PhysRevB.102.094315}. To perform both DMRG and time evolution we use ITensor C++ library \cite{fishman2021itensor}. }
for several atoms taking into account periodic and open boundary conditions. 

\emph{Model.-}
We consider $N$ fermionic ultra-cold atoms, each in two internal states $\ket{\uparrow}$ and $\ket{\downarrow}$ corresponding to a spin-1/2 degree of freedom, loaded into a one-dimensional optical lattice potential of $M$ sites. 
The atoms are assumed to occupy the lowest Bloch band, interact through s-wave collisions, and hence can be described by the FHM. 
In addition, we include a laser driving term which induces a position-dependent spin-flip coupling (SFC) between atomic spin states.
The Hamiltonian for such a system reads
\begin{align}
    \hat{H} & = \sum_{j=1}^{M'} \hat{H}^{\text{tunnel}}_j  +
    \sum_{j=1}^M \hat{H}^{\text{int}}_j +
    \sum_{j=1}^M \hat{H}^{\text{L}}_j,\label{eq:FHM}\\
    \hat{H}^\text{tunnel}_j & = -J \sum_{s= \uparrow,\downarrow}
   (\hat{a}^\dagger_{j,s} \hat{a}_{j+1,s} + \text{h.c.}), \\
   \hat{H}^{\text{int}}_j & = U \hat{n}_{j, \uparrow}\hat{n}_{j, \downarrow}, \\  
    \hat{H}^{\text{L}}_j & = \frac{\hbar\Omega}{2} (e^{i  \phi j } \hat{a}^\dagger_{j,\uparrow}\hat{a}_{j,\downarrow} + e^{-i  \phi j }\hat{a}^\dagger_{j,\downarrow}\hat{a}_{j,\uparrow}),
\end{align}
where the fermionic operators $\hat{a}_{j, s}$ annihilate an atom in the $j$th lattice site in the state $s\in \{\uparrow,\downarrow\}$, and  $\hat{n}_{j,s}=\hat{a}^\dagger_{j, s}\hat{a}_{j,s}$ is the corresponding operator of the number of atoms. The upper limit of sumation is $M' = M$ under periodic boundary conditions (PBC), or $M' = M-1$ under the open ones (OBC). 
The terms $\hat{H}^{\text{tunnel}}_j$ and $\hat{H}^{\text{int}}_j$ 
describe the FHM including on-site nearest-neighbor tunneling of fermions with rate $J$ and on-site repulsion of strength $U$. 
The term $\hat{H}^{\rm L}_j$ represents the on-site laser coupling, with amplitude $\hbar\Omega$ and position-dependent phase $\phi j$, where $\phi=\pi \cos({\eta})\,  \lambda_{\mathrm{latt}}/\lambda_L $ can be tuned by properly choosing the angle $\eta$ between laser beams producing the optical lattice and the direction of the laser field inducing the SFC.  The wavelength of the latter field $\lambda_L$ can differ from the underlying lattice wavelength $\lambda_{\rm latt}$, as depicted in Fig.~\ref{fig:fig1}. The system has been realized experimentally using the optical clock transition between two electronic orbital states of $^{87}$Sr atoms~\cite{Campbell3987, Bromley_2018}. 
The FHM can also be experimentally simulated using tweezer arrays~\cite{Young2020, https://doi.org/10.48550/arxiv.2110.15398}.

We consider the system in the Mott insulating phase for $U \gg J$, with even $N$ and at half-filling, $M=N$, when double occupancy of a single site is energetically unfavorable. 
The second order processes, obtained by a projection of the Hamiltonian onto the low energy manifold by using the Schrieffer--Wolff transformation~\cite{PhysRevResearch.3.013178,Chao_1977,PhysRevB.18.3453,PhysRev.149.491,BRAVYI20112793,PhysRevResearch.3.013178}, lead to the nearest-neighbor spin exchange (SE) interaction~\cite{Demler2003} corrected by the atom-light coupling term:
\begin{align}\label{eq:Heff}
 	\hat{H}_{\rm spin} & = \sum_{j=1}^{N'} \hat{H}^{\rm SE}_j + \sum_{j=1}^N \hat{H}^{\uparrow\downarrow}_j, \\
 	\hat{H}^{\rm SE}_j & =  
 	  J_\text{SE}\left(
	\hat{S}^x_{j} \hat{S}^x_{j+1} + \hat{S}^y_{j} \hat{S}^y_{j+1} + \hat{S}^z_{j} \hat{S}^z_{j+1} - \frac{1}{4}
 	\right),\\
    \hat{H}^{\uparrow\downarrow}_j & =  J_{\uparrow\downarrow} 
  \left(e^{i \phi j} \hat{S}^+_{j} +  e^{-i \phi j} \hat{S}^-_{j}\right),\label{eq:SFCj}
\end{align}
where the on-site spin operators are
$\hat{S}^{+}_j = \hat{a}^\dagger_{j,\uparrow}\hat{a}_{j,\downarrow}$, $\hat{S}^{-}_j =\hat{a}^\dagger_{j,\downarrow}\hat{a}_{j,\uparrow}$, $\hat{S}^{\pm}_j = \hat{S}^{x}_j \pm i\hat{S}^y_j$ and $\hat{S}^{z}_j = (\hat{n}_{j,\uparrow}-\hat{n}_{j,\downarrow})/2$.
The form of the spin Hamiltonian~(\ref{eq:Heff}), with $J_{\text{SE}} \approx 4 J^2/U$ and $J_{\uparrow\downarrow} \approx \hbar \Omega/2$, is valid when $U \gg \hbar\Omega$. The spin Hamiltonian (\ref{eq:Heff}) can be obtained by making $\hat{H}^{\rm L}_j$ position-independent via a $j$-dependent spin rotation~\cite{PhysRevResearch.3.013178} and returning to the original frame at the end of calculations; see derivation and general form of (\ref{eq:Heff}) in Sec.~I of Supplementary Materials (SM)~\cite{supplementary-materials}.

\emph{Simulation of the OAT model.-}
If there is no atom-light coupling, $\Omega=0$, the spin Hamiltonian reduces to the SE Hamiltonian $\hat{H}_{\rm SE}=\sum_j \hat{H}^{\rm SE}_j$ whose eigenstates include
Dicke states $|N/2, m\rangle\propto \hat{S}_{-}^{N/2 - m} \bigotimes_{j=1}^N \ket{\uparrow}_j$ characterized by zero eigenenergy~\cite{supplementary-materials}. Here, $S=N/2$ is the spin quantum number and $m=-S,-S+1,\cdots,S$ is the spin projection quantum number, i.e. $\hat{S}^2|N/2,m\rangle = N/2(N/2-1)|N/2,m\rangle$ and $\hat{S}_z|N/2,m\rangle = m|N/2,m\rangle$, where $\hat{S}^2 = \hat{S}_x^2+\hat{S}_y^2+\hat{S}_z^2$ and $\hat{S}_w = \sum_{j=1}^N \hat{S}^w_j$
are the collective spin operators for $w=x,y,z,\pm$. Consequently, the initial spin coherent state $|\theta, \varphi \rangle=e ^{-i \hat{S}_z \varphi} e^{-i \hat{S}_y \theta} \bigotimes_{j=1}^N \ket{\uparrow}_j$ does not evolve in time and no squeezing can be generated.
A non-trivial evolution of the initial state appears when $\Omega \ne 0$ due to the atom-light coupling term $\hat{H}_{\uparrow\downarrow}=\sum_j \hat{H}_j^{\uparrow \downarrow}$ . In particular, the action of $\hat{H}_{\uparrow\downarrow}$ onto the Dicke states gives $\hat{H}_{\uparrow\downarrow} |N/2, m\rangle = 
J_{\uparrow \downarrow} c^{-1}_{N/2, m+1} |q; m+1\rangle
-J_{\uparrow \downarrow} c^{-1}_{N/2, m-1} |q; m-1\rangle$,
where 
\begin{equation}\label{eq:SWS}
    |q; m\rangle \equiv \pm c_{N/2,\pm m}\sum_{j=1}^N e^{\pm i q j} \hat{S}^\pm_j |N/2, m\mp 1\rangle
\end{equation}
are spin-wave states~\cite{Swallows_2011} for PBC, with
$c_{N/2, m}=\sqrt{N-1}/\sqrt{\left(\frac{N}{2}-m\right)\left(\frac{N}{2}-m+1\right)}$, $\,\,\,q= 2 \pi n/N$ and $n=\pm 1,\pm 2,\cdots, \pm (N/2-1), N/2$. 
They are eigenstates of the total spin operator and its projection, $\hat{S}^2|q; m\rangle = N/2 (N/2-1)|q; m\rangle$ and $\hat{S}_z|q; m\rangle = m |q; m\rangle$, with $-N/2+1\le m\le N/2-1$. The spin-wave states can be constructed starting with the state for maximal projection, $|q; m=N/2-1
\rangle \equiv \frac{1}{\sqrt{N}} \sum_l \hat{S}_l^- e^{i q j} \bigotimes_{j=1}^N \ket{\uparrow}_j$~\cite{Stancil2009}. Subsequently, applying the spin lowering operator $\hat{S}_-$, the state for any $m$ is given by $|q;m\rangle \propto \hat{S}_-^{N/2-1-m}|q; m=N/2-1\rangle$. The spin-wave states~(\ref{eq:SWS}) are eigenstates of $\hat{H}_{\rm SE}$ with eigenenergies $ E_{q}= J_{\rm SE} \left(1- \cos q \right)$ which do not depend on the spin projection $m$ due to the spherical symmetry of the SE Hamiltonian.

The atom-light interaction thus couples the manifold of the Dicke states $|N/2, m\rangle$ to that of the spin wave states $|q; m\rangle$ with $q=\phi$. The two manifolds, labelled by ${\cal D}_{N/2}$ and ${\cal D}_{N/2-1}$, are separated by the energy gap $E_{\phi}=J_{\rm SE} \left(1- \cos \phi \right)$. 
In the weak atom-light coupling regime, $J_{\uparrow \downarrow} \ll E_{\phi}$, the energy mismatch $E_{\phi}$ suppresses the population transfer between the two manifolds. The effect of the spin wave states on the Dicke states can thus be treated perturbatively. The first non-vanishing correction to an effective Hamiltonian for the  Dicke manifold ${\cal D}_{N/2}$ comes in the second order with respect to $\hat{H}^{\uparrow\downarrow}$, giving
\begin{equation}\label{eq:HSWgeneral}
    \hat{H}_{\rm eff}^{(2)} = \hat{I}_{N/2} \hat{H}^{\uparrow\downarrow} \hat{G}_{N/2-1} \hat{H}^{\uparrow\downarrow}
    \hat{I}_{N/2} ,
\end{equation}
where $\hat{I}_{N/2}=\sum_m |N/2,m \rangle \langle N/2, m|$ is the unit projection operator onto the Dicke manifold, while $\hat{G}_{N/2-1}=\sum_{q\ne 0, m} \frac{|m,q\rangle\langle m, q|}{-E_{q}}$ is an operator which sums projectors onto the ${\cal D}_{N/2-1}$ manifold with the corresponding energy mismatch denominator $- E_{q}$.
The Hamiltonian (\ref{eq:HSWgeneral}) can be expressed in terms of collective spin operators by relating its matrix elements $\langle N/2, m|\hat{H}_{\rm eff}^{(2)} |N/2, m'\rangle$ with the matrix elements of spin operators $\hat{S}_x^2$ and $\hat{S}_z^2$,
as explained in more details in  Sec.~II of SM~\cite{supplementary-materials}. 
The form of $\hat{H}^{(2)}_{\rm eff}$ depends on the value of the phase $\phi$. Specifically, one can distinguish two cases:
(i) For $\phi = \pi$ the effective Hamiltonian reads 
\begin{equation}\label{eq:OAT_x}
    \hat{H}_{\rm eff}^{(2)} = - \hbar\chi_{\pi} \hat{S}_x^2,\,\,\mathrm{with}\,\, \hbar\chi_{\pi} = \frac{1}{2}\frac{ J_{\uparrow\downarrow}^2}{J_\mathrm{SE}}\frac{1}{N-1}\,,
\end{equation}
and (ii) for $\phi =  2 \pi n/N$ with 
$n=\pm 1,\pm 2,\cdots, \pm (N/2-1)$ we obtain
\begin{equation}\label{eq:OAT_z}
    \hat{H}_{\rm eff}^{(2)} = \hbar\chi_\phi \hat{S}_z^2,\,\,\mathrm{with}\,\,\hbar\chi_{\phi} = \frac{1}{2(1-\cos\phi)}\frac{ J_{\uparrow\downarrow}^2}{J_\mathrm{SE}}\frac{1}{N-1}\,.
\end{equation}
In both cases we omit the constant energy term proportional to $\hat{S}^2$. This shows that $\hat{H}_{\rm eff}^{(2)}$ has the form of the OAT model in which the axis of twisting is determined by the value of $\phi$ while the  direction of twisting is clockwise in (\ref{eq:OAT_z}) or counter-clockwise in (\ref{eq:OAT_x}); for numerical demonstrations see Sec.~IV of SM~\cite{supplementary-materials}. The dependence on $\phi$ of the squeezing term will be used to develop a protocol for simulation of the TACT model.  Before that, however, let us discuss the numerical results and a time scale for the best spin-squeezing generation. 

\begin{figure}
\centering
    \includegraphics[width=\linewidth]{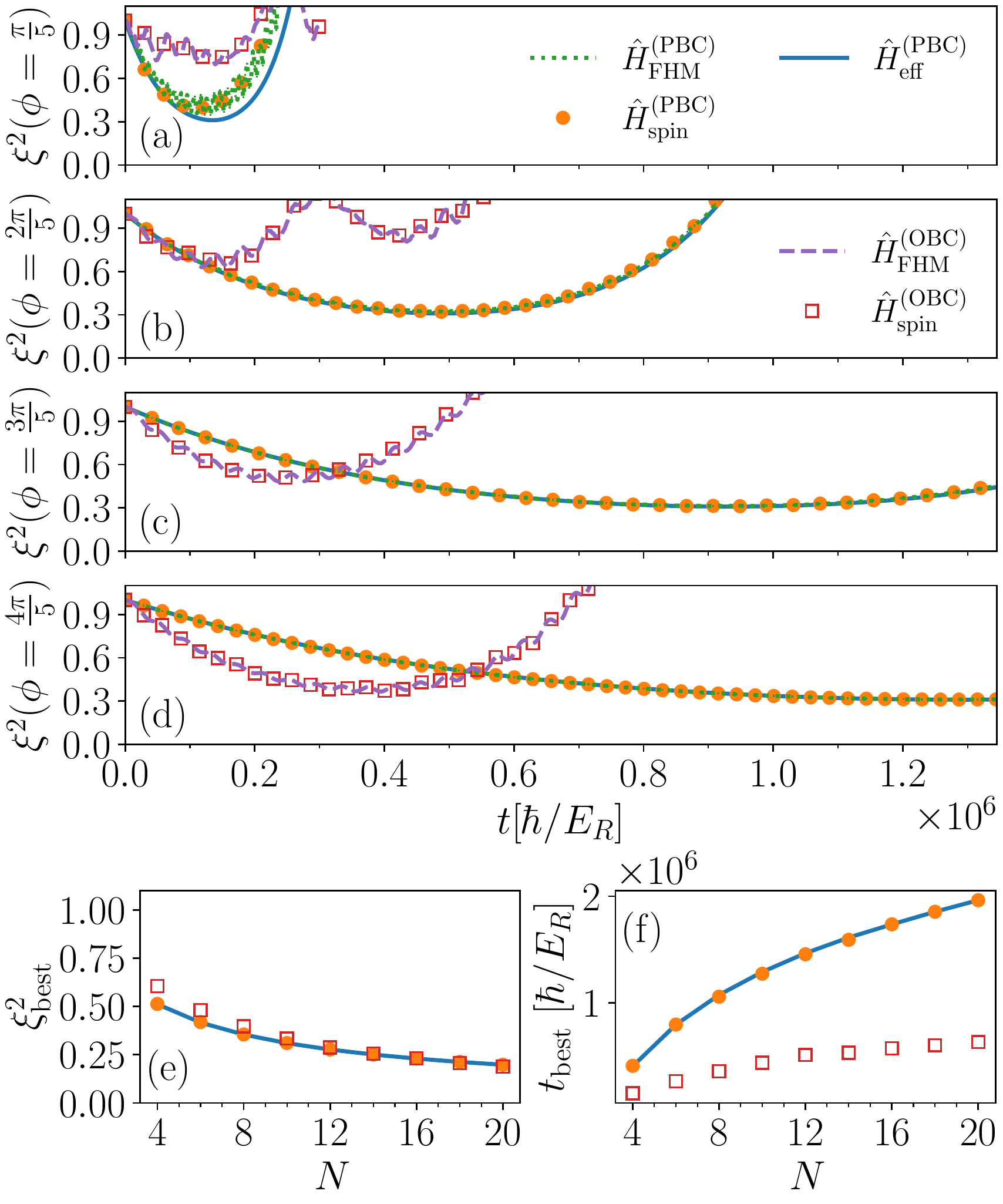}
        \caption{Temporal dependence of the spin squeezing parameter $\xi^2$ for the initial state $|\theta=\pi/2, \varphi=\pi/2\rangle$ evolved with the FHM (\ref{eq:FHM}) (green dotted lines), the spin Hamiltonian (\ref{eq:Heff}) (orange circles) and the effective OAT model (\ref{eq:OAT_x}) (blue lines) under PBC for $N=10$, $J/U=0.04$, $J_{\rm SE}=0.0032E_R$, $J_{\uparrow \downarrow}/J_{\rm SE}=0.04$ (for $U=0.5E_R$, $J=0.02 E_R$ where $E_R=\hbar^2 (2 \pi)^2/(2 m \lambda_{\rm latt}^2)$ is the recoil energy),
        and (a)
        $\phi = \pi/5$, $J_{\uparrow \downarrow}/E_\phi \approx 0.21$, 
        (b) $\phi = 2\pi/5
        $, $J_{\uparrow \downarrow}/E_\phi \approx 0.06$, 
        (c) $\phi = 3 \pi/5$, $J_{\uparrow \downarrow}/E_\phi \approx 0.03$ and 
        (d) $\phi = 4 \pi/5$, $J_{\uparrow \downarrow}/E_\phi \approx 0.02$.
        The results for OBC are shown for the same parameters with the FHM (red squares) and the spin Hamiltonian (violet dashed lines). The panels (e) and (f) represent the best squeezing $\xi^2_\mathrm{best}$ and the best squeezing time $t_\mathrm{best}$, respectively, versus the number of atoms $N$ for the spin (orange circles) and OAT (blue line) models for $\phi = \pi + 2 \pi/N$. Results for the OBC (red squares) are shown for comparison. 
        }
\label{fig:fig2}
\end{figure}

In Fig.~\ref{fig:fig2}(a)-(d) we present the time evolution of the spin squeezing parameter $\xi^2$ for the FHM with the atom-light coupling~(\ref{eq:FHM}) and the spin model (\ref{eq:Heff}). The results are obtained numerically by full many-body calculations, and compared to the solution of the OAT model for various values of the phase $\phi$. The state is spin-squeezed whenever $\xi^2< 1$. The level of the best squeezing (the minimal value of $\xi^2$), as well as the best squeezing time, achieved by the FHM and the spin model agree with OAT as long as the energy gap is large compared to the strength of atom-light coupling $E_{\phi} \gg J_{\uparrow\downarrow}$, 
as required by the validity of the perturbation theory.
The squeezing time diminishes when lowering the value of $\phi$, and it can be further shortened by optimizing parameters $J$ and $U$. The best squeezing and the best squeezing time versus $N$ for PBC (closed circles) and OBC (open squares) are shown in panels (e) and (f), respectively. Note, while the overall level of the best squeezing is similar for PBC and OBC for various values of $N$, the time scale for the best squeezing is shorter for OBC.
In the limit of a large number of atoms, $N \gg 1$, the scaling of the best squeezing time with $N$ is $\chi_{\phi} t_{best} \simeq N^{-2/3}$\cite{PhysRevA.47.5138} implying $E_R t_{best}/\hbar\sim N^{1/3}$ for the fixed value of $\phi$, and this tendency is visible in~Fig.~\ref{fig:fig2}(e).

We have also verified numerically that the unitary evolution governed by the FHM (\ref{eq:FHM}) gives rise to the existence of an Anderson’s tower of states~\cite{PhysRevA.105.022625} corresponding to the Dicke manifold. This strongly indicates that the FHM with the atom-light coupling term features spin squeezing. 
It is also worth noting here that our derivation of Eq.~(\ref{eq:OAT_x}) provides a rigorous mathematical confirmation of the 
quantum simulator arguments used to show the equivalence between the OAT and the XXX model with a staggered field~\cite{PhysRevLett.126.160402}.

\begin{figure}
    \centering
    \includegraphics[width=\linewidth]{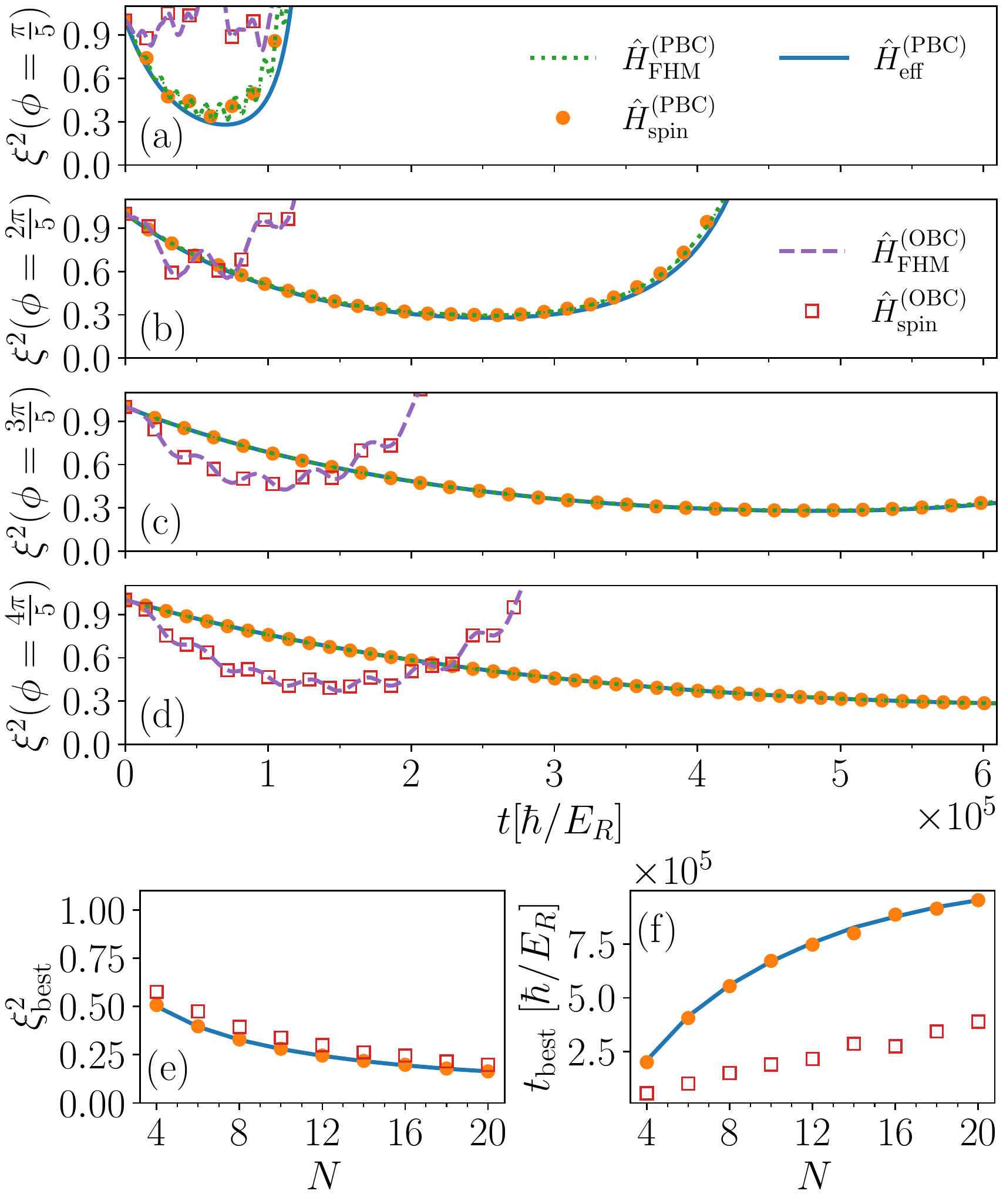}
    \caption{
    Spin squeezing parameter $\xi^2$ in time when the system is coupled by two laser beams to simulate TACT from the initial state $|\theta=\pi/2, \varphi=\pi/2\rangle$. The evolution with the FHM (\ref{eq:FHM}) (marked by green squares), the spin Hamiltonian (\ref{eq:Heff}) (orange circles) and the effective TACT model (\ref{eq:TACT}) (blue line) under PBC is shown for $\phi_0 = \pi$,  $J_{\uparrow\downarrow}^{(1)}/J_{SE} =0.04$ and
    (a) $\phi_1 = \pi/5$ , (b) $\phi_1 = 2\pi/5$, (c) $\phi_1 = 3 \pi/5$ and (d) $\phi_1 = 4 \pi/5$. The results for OBC are shown for the same parameters using the FHM (red squares) and the spin Hamiltonian (violet dashed lines). The best squeezing $\xi^2_\mathrm{best}$ (e) and the best squeezing time $t_\mathrm{best}$ (f) versus the number of atoms $N$ for the spin (orange circles for PBC and red squares for OBC) and TACT (blue line) models for $\phi_1 = \pi + 2 \pi/N$. Other parameters are the same as in Fig.~\ref{fig:fig2}.
    }
\label{fig:scaling_TACT}
\end{figure}

\emph{Simulation of TACT by means of two driving fields.-}
Suppose now that our system is affected by two laser beams producing SFC characterised by two different phases $\phi_{0}=\pi$ and
$\phi_{1}\ne\pi$. Each beam corresponding to the phase $\phi\equiv\phi_{l}$
(with $l=0,\,1$) provides the spin-flip term $\hat{H}^{\uparrow\downarrow}(\phi_{l})$ with the amplitude $J_{\uparrow\downarrow}^{(l)}\approx \hbar \Omega^{(l)}/2$.
The full SFC contribution is then described by the sum of the two terms
$\hat{H}^{\uparrow\downarrow}=\hat{H}^{\uparrow\downarrow}(\phi_{0})+\hat{H}^{\uparrow\downarrow}(\phi_{1})$. In that case the second order effective Hamiltonian given by Eq.(\ref{eq:HSWgeneral}) provides two terms corresponding to contributions by both laser beams  
\begin{equation}
\hat{H}_{\rm eff}^{(2)} =\hbar \chi_{\phi_1}\left( \hat{S}_{z}^{2}-\frac{\chi_{\pi}}{\chi_{\phi_{1}}}\hat{S}_{x}^{2}\right)\,,\label{eq:TACT}
\end{equation}
where we omitted the constant energy shift. 
In general, the resulting model (\ref{eq:TACT}) represents a non-isotropic TACT. However, by taking carefully chosen spin-flip amplitudes $J^{(0)}_{\uparrow\downarrow}= J^{(1)}_{\uparrow\downarrow}/ \sqrt{1-\cos(\phi_{1})}$, one has $\chi_{\pi}=\chi_{\phi_{1}}$, leading to the pure TACT model.

In Fig.~\ref{fig:scaling_TACT}(a)-(d) we show the spin squeezing dynamics with the FHM including the two atom-light coupling fields. The dynamics is seen to successfully simulate the TACT model for PBC. The use of OBC also supports the TACT generation, with a lower level of squeezing but a faster best squeezing time, as illustrated in Fig.~\ref{fig:scaling_TACT}(e) and (f). In the limit of a large number of atoms, the scaling of the best squeezing time for the TACT model is $\chi_{\phi_1} t_{best} \sim N^{-1}\log(2 N)$~\cite{PhysRevA.92.013623} and for fixed $\phi$ gives in our case $t_{best}\sim \log(2 N)$.

\emph{Conclusions.-}
We considered Ramsey-type spectroscopy in the atomic Fermi-Hubbard model with the position-dependent atom-light coupling. We showed that the FHM and the corresponding Heisenberg spin model generate the same dynamics as the OAT and TACT in the weakly coupling regime. Such a regime corresponds to the system in a Mott insulating phase with sufficiently weak atom-light coupling which maintains single-site occupation at half-filling and ensures that the generation of spin squeezing is protected by the energy gap $E_\phi$. Our analytical results are confirmed by the full many-body numerical simulations. 
The scheme is suitable for preparation of spin squeezed, many-body entangled and Bell correlated states~\cite{Panfil2020,Plodzien2022} for state-of-the-art optical lattice clocks using either optical latices~\cite{Bromley_2018} or tweezers arrays~\cite{https://doi.org/10.48550/arxiv.2110.15398}, a key resource for future quantum technologies. 

\section*{ACKNOWLEDGMENTS}
We gratefully acknowledge discussions with Karol Gietka, Bruno Laburthe-Tolra, Alice Sinatra, Ana Maria Rey and Martin Robert-de-Saint-Vincent. This work was supported by DAINA project of the Polish National Science Center DEC-2020/38/L/ST2/00375 and Lithuanian Research Council S-LL-21-3. T.~H.~Y. acknowledge support from the Polish National Science Center DEC-2019/35/O/ST2/01873. M.~P. acknowledges the support of the Polish National Agency for Academic Exchange, the Bekker programme no: PPN/BEK/2020/1/00317. M.~P. also acknowledges support from Agencia Estatal de Investigación (the R\&D project CEX2019-000910-S, funded by MCIN/AEI/10.13039/501100011033, Plan National FIDEUA PID2019-106901GB-I00, FPI), Fundació Privada Cellex, Fundació Mir-Puig, and from Generalitat de Catalunya (AGAUR Grant No. 2017 SGR 1341, CERCA program). E.~W. was partially supported by the NCN Grant No. 2019/32/Z/ST2/00016 through the project MAQS under QuantERA, which has received funding from the European Union’s Horizon 2020 research and innovation program
under Grant Agreement No. 731473. M.~P. acknowledges the computer resources at MareNostrum and the technical support provided by BSC (RES-FI-2022-1-0042). A part of computations were carried out at the Centre of Informatics Tricity Academic Supercomputer \& Network.

\begin{widetext}

\section*{Supplementary Materials}

\section{Derivation of spin Hamiltonian from the FHM with atom-light coupling}\label{app:SWT}

\subsection{Hamiltonian in original and transformed frames}

We consider the Fermi Hubbard (FH) Hamiltonian at half-filling, $M=N$, including the atom-light coupling
\begin{equation}\label{eq:FHM}
    \hat{H} =\hat{H}^\mathrm{tunnel}+\hat{H}^\mathrm{int} + \hat{H}^\mathrm{L},
\end{equation}
with
\begin{align}
    \hat{H}^\mathrm{tunnel} & = -J \sum_j
   \sum_{s= \uparrow,\downarrow}
   (\hat{a}^\dagger_{j,s} \hat{a}_{j+1,s} + \text{h.c.}),
    \\
    \hat{H}^\mathrm{int} & = 
   U \sum_j \hat{n}_{j, \uparrow}\hat{n}_{j, \downarrow}, \\  
    \hat{H}^{\text{L}} & = \frac{\Omega}{2} \sum_j(e^{i  \phi j } \hat{a}^\dagger_{j,\uparrow}\hat{a}_{j,\downarrow} + e^{-i  \phi j }\hat{a}^\dagger_{j,\downarrow}\hat{a}_{j,\uparrow}),
    \label{eq:FHM-3}
\end{align}
where $\hat{n}_{j,\uparrow} = \hat{a}^\dagger_{j,\uparrow}\hat{a}_{j,\uparrow}$ and $\hat{n}_{j,\downarrow} = \hat{a}^\dagger_{j,\downarrow}\hat{a}_{j,\downarrow}$ are the particle number operators for atoms in the spin up and down states at the lattice site $j$. 
For simplicity of notation we set $\hbar=1$ in Eq.~(\ref{eq:FHM-3}) and below. 
We will show how to obtain the full expression for the spin model Hamiltonian under both periodic and open boundary conditions when the interaction exceeds the tunneling, $U\gg J$.
To do this, we used the Schrieffer-Wolf transformation which is described in details in the last section of the Supplementary Materials.
To carry out the analysis we will first diagonalize the atom-light operator $\hat{H}^{\text{L}}$ via introduction of new fermion creation and annihilation operators according to
\begin{align}
	\hat{b}^\dagger_{j,\uparrow} &= 
	\frac{1}{\sqrt{2}} \left( \hat{a}^\dagger_{j,\uparrow} - e^{-i\phi j}\hat{a}^\dagger_{j,\downarrow} \right) \\
	\hat{b}^\dagger_{j,\downarrow} &= 
	\frac{1}{\sqrt{2}} \left( \hat{a}^\dagger_{j,\uparrow} + e^{-i\phi j}\hat{a}^\dagger_{j,\downarrow} \right)\,,
\end{align}	
giving
\begin{equation}
\hat{\mathcal{H}}^{\text{L}} = \Omega \hat{J}_{z} = \frac{\Omega}{2} \sum_j( \hat{b}^\dagger_{j,\uparrow}\hat{b}_{j,\uparrow}-
\hat{b}^\dagger_{j,\downarrow}\hat{b}_{j,\downarrow}  ) \,,
\label{eq:H_L}
\end{equation}
Transition to the new Fermi operators corresponds effectively to the following site-dependent spin rotation
\begin{align}
    \hat{J}^x_{j} &= \hat{S}^z_{j} \label{eq:spintransformation}\\ 
    \hat{J}^y_{j} &= \hat{S}^x_{j} \sin{(\phi j)} + \hat{S}^y_{j} \cos{(\phi j)}\\
    \hat{J}^z_{j} &= -\hat{S}^x_{j} \cos{(\phi j)} + \hat{S}^y_{j} \sin{(\phi j)}\, \label{eq:spintransformation-last},
\end{align}
where
$\hat{S}^x_{j} = \left(\hat{a}^\dagger_{j,\downarrow}\hat{a}_{j,\uparrow}+
\hat{a}^\dagger_{j,\uparrow}\hat{a}_{j,\downarrow}\right)/2$,
$\hat{S}^y_{j} = i \left(\hat{a}^\dagger_{j,\downarrow}\hat{a}_{j,\uparrow}-
\hat{a}^\dagger_{j,\uparrow}\hat{a}_{j,\downarrow}\right)/2$,
and
$\hat{S}^{z}_j = (\hat{a}^\dagger_{j,\uparrow}\hat{a}_{j,\uparrow}-\hat{a}^\dagger_{j,\downarrow}\hat{a}_{j,\downarrow})/2$ are the original spin operators at the $j$th lattice site.

In terms of new operators $\hat{b}^\dagger_{j,\uparrow, \downarrow}$ and $\hat{b}_{j,\uparrow, \downarrow}$ the tunneling and interaction terms comprising the Hamiltonian (\ref{eq:FHM}) read 
\begin{align}
    \hat{\mathcal{H}}^\mathrm{tunnel} =& -\frac{J}{2} \sum_j \left[ 
    (1+e^{-i\phi})(\hat{b}^\dagger_{j,\uparrow} \hat{b}_{j+1,\uparrow} + \hat{b}^\dagger_{j,\downarrow} \hat{b}_{j+1,\downarrow}) 
    +(1-e^{-i\phi})(\hat{b}^\dagger_{j,\uparrow} \hat{b}_{j+1,\downarrow} + \hat{b}^\dagger_{j,\downarrow} \hat{b}_{j+1,\uparrow})
    + \mathrm{h.c.}\right] ,\\ 
    \hat{\mathcal{H}}^\mathrm{int} =& U \sum_j \hat{N}_{j,\uparrow} \hat{N}_{j,\downarrow}, 
\end{align}
with  $\hat{N}_{j,\uparrow} = \hat{b}^\dagger_{j,\uparrow}\hat{b}_{j,\uparrow}$ and $\hat{N}_{j,\downarrow} = \hat{b}^\dagger_{j,\downarrow}\hat{b}_{j,\downarrow}$ being the corresponding particle number operators in the spin up and down states with respect to the rotated basis. 

To proceed with the calculations it is convenient to split the Hamiltonian (\ref{eq:FHM}) into a sum of two-sites Hubbard Hamiltonian~\cite{Chao_1977}, as 
\begin{align}
    \hat{\mathcal{H}}^{\mathrm{(PBC)}} &= \sum^N_{\langle i, j \rangle}  \hat{\mathcal{H}}_{i,j} , \label{eq:H_rot_periodic}\\
    \hat{\mathcal{H}}^{\mathrm{(OBC)}}     &= \sum^{N-1}_{\langle i, j \rangle}  \hat{\mathcal{H}}_{i,j} + \hat{\mathcal{H}}^{\mathrm{(OBC)}}_{N, 1} \label{eq:H_rot_open},
\end{align}
where the summation $\langle i, j \rangle$ runs over all the nearest-neighbour pairs, the superscripts $^{\mathrm{(PBC)}}$ and $^{\mathrm{(OBC)}}$ refer, respectively, to periodic boundary conditions (PBC) and open boundary conditions (OBC) and $N$ is the number of lattice sites. The bulk pair term $\hat{\mathcal{H}}_{j,j+1}$ is given by 
\begin{equation}\label{eq:FH_rotated_pairs}
\begin{split}
	\hat{\mathcal{H}}_{j,j+1} =& -\frac{J}{2} \left[ 
                (1+e^{-i\phi})(\hat{b}^\dagger_{j,\uparrow} \hat{b}_{j+1,\uparrow} + \hat{b}^\dagger_{j,\downarrow} \hat{b}_{j+1,\downarrow}) 
                +(1-e^{-i\phi})(\hat{b}^\dagger_{j,\uparrow} \hat{b}_{j+1,\downarrow} + \hat{b}^\dagger_{j,\downarrow} \hat{b}_{j+1,\uparrow})  
                + \mathrm{h.c.} \right] \\ 
		  &+ U \left( \hat{N}_{j,\uparrow} \hat{N}_{j,\downarrow} 
		  + \hat{N}_{j+1,\uparrow} \hat{N}_{j+1,\downarrow}  \right) \\ 
		  &+ \Omega \left( \hat{J}^z_{j} + \hat{J}^z_{j+1}   \right) ,
\end{split}
\end{equation}
For OBC there is no contribution by the intersite tunneling in the last term $\hat{\mathcal{H}}^{\mathrm{(OBC)}}_{N, 1}$ involving a pair of the sites $N$ and $1$, which is expressed as
\begin{equation} \label{eq:H_M-1_open}
	\hat{\mathcal{H}}^{\mathrm{(OBC)}}_{N, 1} = 
		   U \left( \hat{N}_{1,\uparrow} \hat{N}_{1,\downarrow} 
		  + \hat{N}_{N,\uparrow} \hat{N}_{N,\downarrow}  \right) \\ 
		  + \Omega \left( \hat{J}^z_{1} + \hat{J}^z_{N}   \right).
\end{equation}

\subsection{Effective spin Hamiltonian}

For each pair of the lattice sites $j$ and $j+1$ coupled by the intersite tunneling we use the basis spin states for transformed spin operators including the following two subsets: (i) four states describing single occupied lattice sites 
$|n_1\rangle = |\uparrow\rangle_j|\uparrow\rangle_{j+1}$,
$|n_2\rangle =|\uparrow\rangle_j|\downarrow\rangle_{j+1}$, 
$|n_3\rangle =|\downarrow\rangle_j|\uparrow\rangle_{j+1}$, 
$|n_4\rangle =|\downarrow\rangle_j|\downarrow\rangle_{j+1}$ 
and (ii) two states corresponding to double occupied sites 
$|\xi_{1}\rangle = |\uparrow \downarrow\rangle_j|0\rangle_{j+1}$
$|\xi_{2}\rangle = |0\rangle_j|\uparrow \downarrow\rangle_{j+1}$, where the $j$-dependence of the states $|n_{l}\rangle$ and $|\xi_{p}\rangle$ is kept implicit. Both $\hat{\mathcal{H}}^\mathrm{int}$  and $\hat{\mathcal{H}}^\mathrm{L}$ are diagonal in these basis states:
\begin{align}
    (\hat{\mathcal{H}}^{\rm int} + \hat{\mathcal{H}}^{\rm L}) |\xi_p\rangle = E_{\xi_p} |\xi_p\rangle \\
    (\hat{\mathcal{H}}^{\rm int} + \hat{\mathcal{H}}^{\rm L}) |n_l\rangle = E_{n_l} |n_l\rangle,
\end{align}
with $E_{\xi_{1,2}}=U$, $E_{n_1} = -E_{n_4} =\Omega$ and $E_{n_{2,3}}=0$. 
Then, if $J \ll U - \Omega$, we can treat the tunnelling term as a perturbation to the single occupied states $\{ |n_{l=1,2,3,4}\rangle \}$. To fulfill this condition we will assume $J\ll U$ and $\Omega\ll U$ for convenience in the following calculations.
Applying the Schrieffer-Wolff (SW) transformation, one arrives at the following matrix elements of the effective Hamiltonian acting on the manifold of single occupied lattice sites $|n_{l}\rangle $ with $l=1,2,3,4$: 
\begin{equation}
H^\mathrm{spin}_{n_l n_{l'}}= E_{n_l}\delta_{l , l'}+\frac{1}{2}\sum_{p=1}^2\left(\frac{V_{n_l \xi_p}V_{\xi_p n_{l'}}}{E_{n_l}-E_{\xi_p}} + \frac{V_{n_l \xi_p} V_{\xi_p n_{l'}}}{E_{n_{l'}} - E_{\xi_p}}\right)\,,
\end{equation}
with $V_{n_l \xi_p} = \langle \xi_p | \hat{\mathcal{H}}^\mathrm{tunnel} | n_l \rangle $, where the second order processes are mediated by virtual transitions to the double occupied states $|\xi_{p}\rangle $ with $p=1,2$.
Such a pairwise Hamiltonian can be represented in the matrix form as 

\begin{equation} 
    H^\mathrm{spin}= 
    \begin{pmatrix} 
        \Omega +\frac{J^2}{U-\Omega} (\cos \phi -1) & i\frac{J^2}{2U}\frac{2U-\Omega}{U-\Omega}\sin\phi & -i\frac{J^2}{2U}\frac{2U-\Omega}{U-\Omega}\sin\phi & -\frac{J^2U}{U^2 -\Omega^2} (\cos \phi -1)\\ 
        -i\frac{J^2}{2U}\frac{2U-\Omega}{U-\Omega}\sin\phi & -\frac{J^2}{U} (\cos \phi +1) & \frac{J^2}{U} (\cos \phi +1) & i\frac{J^2}{2U}\frac{2U+\Omega}{U+\Omega} \sin\phi \\ 
        i\frac{J^2}{2U}\frac{2U-\Omega}{U-\Omega}\sin\phi & \frac{J^2}{U} (\cos \phi +1) & -\frac{J^2}{U} (\cos \phi +1) & -i\frac{J^2}{2U}\frac{2U+\Omega}{U+\Omega} \sin\phi \\ 
        -\frac{J^2U}{U^2 -\Omega^2} (\cos \phi -1) & -i\frac{J^2}{2U}\frac{2U+\Omega}{U+\Omega} \sin\phi & i\frac{J^2}{2U}\frac{2U+\Omega}{U+\Omega} \sin\phi & -\Omega +\frac{J^2}{U+\Omega} (\cos \phi -1)\,.\\ 
    \end{pmatrix} 
\end{equation}

Thus one arrives at the following operator-level form for the contribution to the spin Hamiltonian by the selected pair of lattice sites $j$ and $j+1$:
\begin{equation}\label{eq:SW_rotated_pairs}
\begin{split}
    \hat{\mathcal{H}}^{\text{spin}}_{(j,j+1)}
    =
    & \tilde{J}_{SE}
    \left( \hat{J}^x_{j}\hat{J}^x_{j+1} -\frac{1}{4} \hat{N}_j \hat{N}_{j+1}\right)
    +\tilde{J}_{yz} 
    \left(\hat{J}^y_{j}\hat{J}^y_{j+1}  + \hat{J}^z_{j}\hat{J}^z_{j+1} \right)+\tilde{J}_{DM}
    \left(\hat{J}^y_{j}\hat{J}^z_{j+1}  - \hat{J}^z_{j}\hat{J}^y_{j+1}\right)\\
    &+ \frac{\tilde{J}_{\uparrow\downarrow}}{2}
    \left( \hat{N}_j \hat{J}^z_{j+1} +  \hat{J}^z_{j} \hat{N}_{j+1} \right) 
     +\tilde{J}_{\Delta y}
    \left(\hat{J}^y_{j} \hat{N}_{j+1} - \hat{N}_j \hat{J}^y_{j+1}\right)
    ,
\end{split}
\end{equation}

where

\begin{align}
    \tilde{J}_{SE}  =& \frac{4 J^2}{U(U^2-\Omega^2)}\left(U^2 - \Omega^2\cos^2\frac{\phi}{2}\right),\\
    \tilde{J}_{yz}  =& \frac{4 J^2}{U(U^2-\Omega^2)}\left(U^2\cos\phi - \Omega^2\cos^2\frac{\phi}{2}\right),\\
    \tilde{J}_{DM}  =& \frac{4 J^2}{U(U^2-\Omega^2)}(2U^2 - \Omega^2)\sin\phi,\\
	\tilde{J}_{\uparrow\downarrow} =& 
	\Omega - \Delta\Omega = \Omega - \frac{4 J^2\Omega}{U^2-\Omega^2}\sin^2\frac{\phi}{2},\\
    \tilde{J}_{\Delta y}  =& \frac{ J^2 \Omega}{U^2-\Omega^2}\sin \phi.
\end{align}
The spin Hamiltonian given by Eq.~(\ref{eq:SW_rotated_pairs}) contains the spin-exchange (SE), Dzyaloshinskii-Moriya (DM) terms, as well as the atom-light coupling term with a modified coefficient $\tilde{J}_{\uparrow\downarrow}$. Note that the DM term appears here to a large extent due to the transformation involving the site-dependent spin rotation, so this term characerized by the strength $ \tilde{J}_{DM}$ does not vanish when the Rabi frequency $\Omega$ of the atom light coupling goes to zero. As we will see next, this is not the case  when the Hamiltonian is expressed in terms of the original spin operators after returning to the original frame. 

\subsection{Returning to original frame}
\subsubsection{Periodic boundary conditions (PBC)}

As indicated in Eq.~(\ref{eq:H_rot_periodic}) for PBC, the transformed frame spin Hamiltonian involves summation over all $j$. 
Calling on Eq.~(\ref{eq:SW_rotated_pairs}) one thus has
\begin{equation}\label{eq:H_rot_periodic_full}
    \hat{\mathcal{H}}^{\mathrm{(PBC)}}_\text{spin}
    = 
    \tilde{J}_\mathrm{SE} \sum_{j=1}^N
    \left( \hat{J}^x_{j}\hat{J}^x_{j+1} -\frac{1}{4}\right)
    +\tilde{J}_{yz} \sum_{j=1}^N
    \left(\hat{J}^y_{j}\hat{J}^y_{j+1}  + \hat{J}^z_{j}\hat{J}^z_{j+1} \right)
    +\tilde{J}_\mathrm{DM} \sum_{j=1}^N
    \left(\hat{J}^y_{j}\hat{J}^z_{j+1}  - \hat{J}^z_{j}\hat{J}^y_{j+1}\right) 
    + \tilde{J}_{\uparrow\downarrow}\hat{J}_{z} .
\end{equation}
Returning back into the original spin operators via Eqs.~(\ref{eq:spintransformation})-(\ref{eq:spintransformation-last}) and combining constants, the spin Hamiltonian becomes
\begin{equation}\label{eq:SW-original} 
\begin{split} 
    \hat{H}^{\mathrm{(PBC)}}_\mathrm{spin}=& 
    J_\mathrm{SE} 
    \sum_{j=1}^N \left(\hat{S}^x_{j}\hat{S}^x_{j+1} +\hat{S}^y_{j}\hat{S}^y_{j+1}
    +\hat{S}^z_{j}\hat{S}^z_{j+1} -\frac{1}{4} \right)\\ 
    &
    +J_\mathrm{DM}\sum_{j=1}^N \left(\hat{S}^y_{j}\hat{S}^x_{j+1} -\hat{S}^x_{j}\hat{S}^y_{j+1}  \right)
    + J_{\uparrow\downarrow} \sum_{j=1}^N \left( 
    -\cos\phi j \hat{S}^x_{j} 
    + \sin\phi j \hat{S}^y_{j}\right)   ,
\end{split} 
\end{equation} 
where
\begin{align} 
    J_\mathrm{SE}  =& \frac{4 J^2}{U(U^2-\Omega^2)}(U^2 - \Omega^2\cos^2\frac{\phi}{2})\,,\\ 
    J_\mathrm{DM}  =& \frac{2 J^2}{U(U^2-\Omega^2)}\Omega^2  \sin{\phi} \, ,\\
    J_{\uparrow\downarrow} =& \left( \Omega - \Delta\Omega \right)  
    = \Omega\left(1 - \frac{4 J^2}{U^2-\Omega^2}\sin^2\frac{\phi}{2} \right) \, .
\end{align} 
Note that the DM term appears now only when the atom light coupling is present, the coupling strength $J_\mathrm{DM}$ going to zero as $\Omega \rightarrow 0$. 
Furthermore, the DM term can be omitted as it is very small compared with the other terms. 
In fact, since  $J \ll U$ and $\Omega \ll U$, then 
\begin{alignat}{6}
	\frac{2J_\mathrm{DM}}{J_\mathrm{SE}} 
	    &= \frac{\Omega^2}{U^2-\Omega^2\cos^2\frac{\phi}{2}}\sin{\phi} 
        &\le
        & \qquad \frac{\Omega^2}{U^2-\Omega^2} 
        &\approx \left(\frac{\Omega}{U}\right)^2 
        &\ll 1
        , \label{eq:p_r1}\\
	\frac{J_\mathrm{DM}}{J_{\uparrow\downarrow}} 
	    &= \frac{\Omega}{U}\frac{2 J^2 \sin\phi}{ U^2-\Omega^2-{2J^2}(1-\cos\phi)}
	    &\le
	    &\frac{\Omega}{U}\frac{2 J^2}{ U^2-\Omega^2-{2J^2}} 
	    &\approx 2\frac{\Omega}{U}\left(\frac{J}{U}\right)^2 
	    &\ll 1
	    .\label{eq:p_r2}
\end{alignat}

Using these conditions for the coefficients, the Hamiltonian simplifies to
\begin{equation}
    \hat{H}^{\mathrm{(PBC)}}_\mathrm{spin}\approx J_\mathrm{SE} 
    \sum_j\left(\hat{S}^x_{j}\hat{S}^x_{j+1} +\hat{S}^y_{j}\hat{S}^y_{j+1}
    +\hat{S}^z_{j}\hat{S}^z_{j+1} -\frac{1}{4} \right)
    +
    J_{\uparrow\downarrow} \sum_j \left( 
    -\cos\phi j \hat{S}^x_{j} 
    + \sin\phi j \hat{S}^y_{j}\right)
,\end{equation} 
with $J_{\uparrow\downarrow} \approx \Omega$ and $J_\mathrm{SE} \approx 4 J^2 / U$.

\subsubsection{Open boundary conditions (OBC)}

Consider now the effective spin Hamiltonian for single occupied lattice sites under the OBC.
In that case, one can still apply the SW approach leading to the spin Hamiltonian $\hat{\mathcal{H}}^{\text{spin}}_{(j,j+1)}$ given by Eq.~(\ref{eq:SW_rotated_pairs}) for all pairs with  $j\le N-1$. 
Since there is no tunneling between boundary lattice sites with $j=N$ and $j=1$, no second order SW correction appears for that remaining pair. There is
only a zero-order contribution 
$\Omega ( \hat{J}^z_{1} + \hat{J}^z_{N} )$
corresponding to the atom-light coupling
in Eq.~(\ref{eq:H_M-1_open}) for $\hat{\mathcal{H}}^{\mathrm{(OBC)}}_{N, 1}$.
By putting together all pair contributions, the Hamiltonian takes the form 
\begin{align*}
    \hat{\mathcal{H}}^{\mathrm{(OBC)}}_\text{spin}
	=& 
    \left( \Omega - \Delta\Omega \right) \sum_{j=1}^{N} \hat{J}^z_{j}
    + \frac{1}{2}\left(\Omega + {\Delta\Omega}\right)\left( \hat{J}^z_{N}  + \hat{J}^z_{1} \right)
    +\tilde{J}_{\Delta y}
    \left( \hat{J}^y_{1} - \hat{J}^y_{N} \right)\\
    &  +\sum_{j=1}^{N-1} \bigg( \tilde{J}_\mathrm{SE}
    \left( \hat{J}^x_{j}\hat{J}^x_{j+1} - \frac{1}{4}\right)
    +\tilde{J}_{yz} 
    \left(\hat{J}^y_{j}\hat{J}^y_{j+1}  + \hat{J}^z_{j}\hat{J}^z_{j+1} \right) 
     +\tilde{J}_\mathrm{DM}
    \left(\hat{J}^y_{j}\hat{J}^z_{j+1}  - \hat{J}^z_{j}\hat{J}^y_{j+1}\right)
    \bigg)\,.\\
\end{align*}
In terms of the original spin operators the Hamiltonian takes the form 
\begin{equation}\label{eq:SW-original_open}
\begin{split}
    \hat{H}^{\mathrm{(OBC)}}_\text{spin}=&
    \hat{H}^{\mathrm{(PBC)}}_\text{spin}
    + i\delta\Omega \Big(
	    e^{i\phi (N+\frac{1}{2})}\hat{S}^+_{N}
	    -e^{-i\phi (N+\frac{1}{2})}\hat{S}^-_{N}
	    -\left(e^{i\phi (1-\frac{1}{2})}\hat{S}^+_{1}
	    -e^{-i\phi (1-\frac{1}{2})}\hat{S}^-_{1}\right)
    \Big) \\
    & -\Big(
    \frac{1}{2}\left( J_\mathrm{SE} + i J_\mathrm{DM} \right) \hat{S}^+_{N}\hat{S}^-_{1} 
    +\frac{1}{2}\left( J_\mathrm{SE} - i J_\mathrm{DM} \right) \hat{S}^-_{N}\hat{S}^+_{1}
    +J_\mathrm{SE}\left( \hat{S}^z_{N}\hat{S}^z_{1} -\frac{1}{4}   \right)
    \Big)
, \end{split}
\end{equation}
where
\begin{align}
    \delta \Omega =& \frac{J^2 \Omega}{U^2-\Omega^2}\sin\frac{\phi}{2}.
\end{align}

Since $J_\mathrm{SE} \gg \delta\Omega, J_\mathrm{DM}$, we can ignore those small terms or treat their effect 
on $\hat{H}_\mathrm{SE}$ as a perturbation.
This can be seen from a simple inspection keeping also in mind our assumption $J \ll U; \Omega \ll U$:
\begin{alignat}{3}
	\frac{\delta\Omega}{J_\mathrm{SE}} 
	    &= \frac{\Omega U}{4\left( U^2-\Omega^2\cos^2\frac{\phi}{2}\right)}\sin\frac{\phi}{2} 
	    &\le \frac{\Omega U}{4(U^2-\Omega^2)}
	    \approx \frac{1}{4}\frac{\Omega}{U}
	    \ll 1
	,\label{eq:o_r2}
\end{alignat}
and we already showed that $J_\mathrm{DM}/ J_\mathrm{SE}\ll 1$ in Eq.~(\ref{eq:p_r1}).

Notice that these terms are also much smaller than the main perturbation $J_{\uparrow\downarrow}$ in our range of parameters since $\delta\Omega / J_{\uparrow\downarrow} \approx (J/U)^2 \ll 1$ and we already previously showed that $J_\mathrm{DM}/J_{\uparrow\downarrow} \ll 1$.
Moreover, $J_{\uparrow\downarrow}$ appears in the bulk terms as well, making $\delta \Omega$ even less relevant for large $N$.

We can thus safely assume that $J_\mathrm{DM}$ and the boundary terms do not affect the overall evolution of the system.
Another way to see this is to simply compare the squeezing parameter when neglecting these terms or not, as illustrated in Fig.~(\ref{fig:obc_neglect}) where numerical results support our conclusions about $\delta\Omega$ even for quite small values of $N$. Notice that for $\phi = \pi$,  $J_\mathrm{DM}=0$; so we are only comparing the influence of $\delta\Omega$. 

\begin{figure}[htpb]
	\centering
	\includegraphics[width=0.8\textwidth]{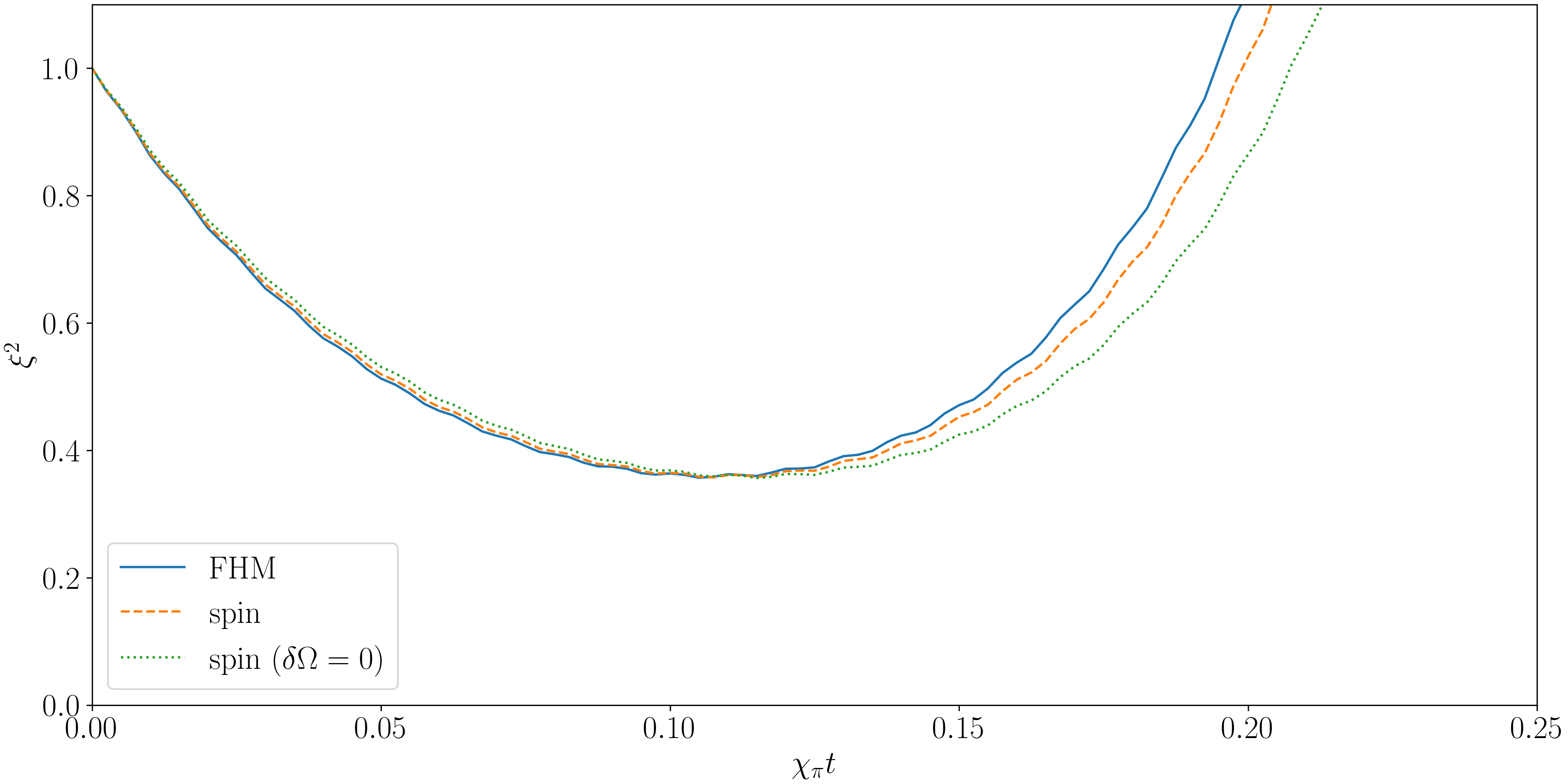}
	\caption{Spin squeezing under OBC with parameters $N=M=8, U/E_R = 1, J/E_R = 0.1, \Omega=10^{-3}, \phi = \pi$. Initial state $\ket{\theta=\frac{\pi}{2}, \phi = \frac{\pi}{2}}$. FHM model is shown by solid blue line while the spin model is marked with the dashed orange line for $\delta\Omega\ne 0$, and by dotted green for $\delta\Omega = 0$.
	}
	\label{fig:obc_neglect}
\end{figure}

In this way, the Hamiltonian may be simplified to the following expression for OBC:
\begin{equation}\label{eq:SW-original_open_simple}
    \hat{H}^{\mathrm{(OBC)}}_\text{spin}
    \approx \hat{H}^{\mathrm{(PBC)}}_\text{spin} - J_\mathrm{SE} \left(\hat{S}^x_{N}\hat{S}^x_{1} 
    +\hat{S}^y_{N}\hat{S}^y_{1}
    +\hat{S}^z_{N}\hat{S}^z_{1} -\frac{1}{4}   \right),
\end{equation}
with $J_\mathrm{SE}  \approx 4 J^2 / U$.

\section{Relating $\hat{H}_{\rm eff}^{(2)}$ to spin-operators}
\label{app:matrix}

Let us consider matrix elements of the spin Hamiltonian $\hat{H}_{\rm eff}^{(2)} $ given by Eq.~(9) in the main text. Its non-zero matrix elements $H_{m, m'}=\langle N/2, m|\hat{H}_{\rm eff}^{(2)} |N/2, m'\rangle$ are 
\begin{eqnarray}
H_{m,m} &=& -\frac{J_{\uparrow\downarrow}^2}{E_{q}}
\left( c^{-2}_{N/2, m} +  c^{-2}_{N/2, -m}\right)  
\label{H_m,m}\\
H_{m,m+2} &=& \frac{J_{\uparrow\downarrow}^2}{E_{q}} c^{-1}_{N/2, -(m+1)} c^{-1}_{N/2, m+1} \delta_{q, -\pi}
\label{H_m,m-plus-2}\\
H_{m,m-2} &=& \frac{J_{\uparrow\downarrow}^2}{E_{q}} c^{-1}_{N/2, m-1} c^{-1}_{N/2, -(m-1)} \delta_{q, -\pi}.
\label{H_m,m-minus-2}
\end{eqnarray}
with $c_{N/2, m}=\sqrt{N-1}/\sqrt{\left(\frac{N}{2}-m\right)\left(\frac{N}{2}-m+1\right)}$. Note that off-diagonal elements $H_{m,m\pm2}$ are non-zero only if $q = \phi = \pi$, 
while diagonal elements are non-zero for all possible values of $q=\phi\ne 0$. 

We will relate the matrix elements of $\hat{H}_{\rm eff}^{(2)} $ to those of $\hat{S}_x^2$ and $\hat{S}_z^2$, and show the validity of Eqs.(10) and (11) of the main text. The spin operators of $\hat{S}_z^2$ and $\hat{S}^2$ are diagonal in the basis of the Dicke states with the matrix elements given by
\begin{align}
\langle N/2, m|  \hat{S}_z^2| N/2, m\rangle &= m^2 ,\\
\langle N/2, m|  \hat{S}^2| N/2, m\rangle &= \frac{N}{2} \left(\frac{N}{2} + 1\right) .
\end{align}
Comparing these two equations with Eqs.~(\ref{H_m,m})-(\ref{H_m,m-minus-2}), for $\phi\ne\pi$  one arrives at the effective OAT model along the $z$ axis, $\hat{H}_{\rm eff}^{(2)} = \chi_\phi \hat{S}_z^2 $, proving Eq.(11) of the main text.

Consider now the square of the spin operator $\hat{S}_x=\left( \hat{S}_{+}+\hat{S}_{-}\right)/2$:
\begin{equation}
\hat{S}_x^2 = \left( \hat{S}_+\hat{S}_- + \hat{S}_-\hat{S}_+ 
+ \hat{S}_+^2 + \hat{S}_-^2
\right) / 4\,.
\label{S_x^2}
\end{equation}
The action of spin raising and lowering operators on the Dicke states is: $\hat{S}_{\pm} |{S,m}\rangle=A_{\pm}^{S,m}|{S,m\pm1}\rangle$, with $A_{\pm}^{S,m}=\sqrt{(S\mp m)(S\pm m+1)}$. Therefore the relevant matrix elements are
\begin{align}
\langle N/2, m| \left( \hat{S}_+\hat{S}_- + \hat{S}_-\hat{S}_+ \right)| N/2, m\rangle &=  2\left( \frac{N^2}{4} - m^2 + \frac{N}{2}\right) , \label{eq:45}\\
\langle N/2, m| \hat{S}_-^2| N/2, m+2\rangle &= \sqrt{\left(\frac{N}{2}+m+2\right)\left(\frac{N}{2}-m-1\right)\left(\frac{N}{2}+m+1\right)\left(\frac{N}{2}-m\right)} \nonumber \\
&=  (N-1) c_{N/2, m+1}^{-1} c_{N/2, -(m+1)}^{-1} ,  \label{eq:46} \\
\langle N/2, m| \hat{S}_+^2| N/2, m-2\rangle &=
\sqrt{\left(\frac{N}{2}+m\right)\left(\frac{N}{2}-m+1\right)\left(\frac{N}{2}+m-1\right)\left(\frac{N}{2}-m+2\right)} \nonumber \\ 
&=(N-1) c_{N/2, m-1}^{-1} c_{N/2, -(m-1)}^{-1} ,  \label{eq:47}
\end{align}
Relating the matrix elements of $\hat{S}_x^2$ given by Eqs. (\ref{S_x^2})-(\ref{eq:47})
with matrix elements of the effective Hamiltonian of  Eqs.~(\ref{H_m,m})-(\ref{H_m,m-minus-2}) corresponding to $q=\phi=\pi$, one arrives at Eq.(10) of the main text, i.e. 
$\hat{H}_{\rm eff}^{(2)} = \chi_\pi \hat{S}_x^2 $.

Note, our results for simulation of OAT by the spin model are also valid for isotropic spin-exchange models with longer range interactions, $\hat{H}^{\rm{SE}} = \sum_{j, j'}  \sum_{\sigma=x,y,z} K_{j, j'} \hat{S}^\sigma_j \hat{S}^\sigma_{j'}  - 1/4$  when $J_{\rm SE}$ is replaced by $\sum_{j, j'} K_{j,j'}$ in (10)-(11) of main text, with $K_{j,j'}$ being a real and symmetric matrix.

\section{Spin squeezing parameter and initial state}\label{app:spinsqueezingparameter}

We quantify the level of squeezing by the spin squeezing parameter
\begin{equation}\label{eq:ssqparameter}
\xi^2 = \frac{N (\Delta S_{\perp})_{\rm  min}^2}{\langle S \rangle^2}
\end{equation}
where the length of the mean collective spin is $\langle S \rangle$ and the minimal variance of the collective spin orthogonally
to its direction is $(\Delta S_{\perp})_{\rm  min}^2$. The collective spin operators $\hat{S}_\gamma = \sum_j \hat{S}^\gamma_j$ with $\gamma=x,y,z,\pm$ give rise to the description of squeezing dynamics from the spin coherent state
\begin{equation}
|\theta, \varphi \rangle = e^{-i \hat{S}_z\varphi} e^{-i \hat{S}_y \theta} \bigotimes_{j=1}^N \ket{\uparrow}_j  
\end{equation}
initialized in the Dicke manifold of the total spin $S=N/2$.
In general, the spin-coherent state can be expressed in the spin basis, $ |\theta, \varphi \rangle = \sum_{m=-S}^S \alpha_m|S=N/2, m \rangle$, where $\alpha_m=\sqrt{\binom{2 S}{S+m}} \cos^{S+m}(\theta/2) \sin^{S-m}(\theta/2) e^{i(S-m)\varphi}$ are coefficients of decomposition. 
Here, $S$ and $m$ label the eigenvalues of the collective spin operators $\hat{S}^2 = \hat{S}_x^2+\hat{S}_y^2+\hat{S}_z^2$ and $\hat{S}_z$, respectively, with eigenvalues $S(S + 1)$ for $S\in (N/2, N/2-1,\cdots )$ and $m\in (-S, -S +1,\cdots, S)$. 

Note, the initial spin coherent state $|\theta, \varphi \rangle$ is an eigenstate of the exchange Hamiltonian $\hat{H}_{SE}$. It is because Dicke states $|S=N/2, m\rangle $ are eigenstates of $\hat{H}_{SE}$ with zero eigenvalue. 
To see this, one needs to calculate first the eigenvalue of $\hat{H}_{SE}$ for $m=N/2$,  obtaining $\hat{H}_{SE}|N/2,N/2\rangle = 0|N/2,N/2\rangle$, where $|N/2,N/2\rangle = \bigotimes_{j=1}^N \ket{\uparrow}_j$. Next, one needs to use $|{S, m}\rangle\propto \hat{S}_-^{S-m} |S, S\rangle$ to obtain the Dicke state for any $m$. Finally, by using the commutation relations $[\hat{S}_-, \hat{H}_{SE}]=0$ one finds that $\hat{H}_{SE} |S=N/2, m\rangle = 0 |S=N/2, m\rangle$.

\section{Dependence of the twisting axis on $\phi$ for the simulation of OAT model}
\label{app:twisting}

\begin{figure}
\centering
    \includegraphics[width=\linewidth]{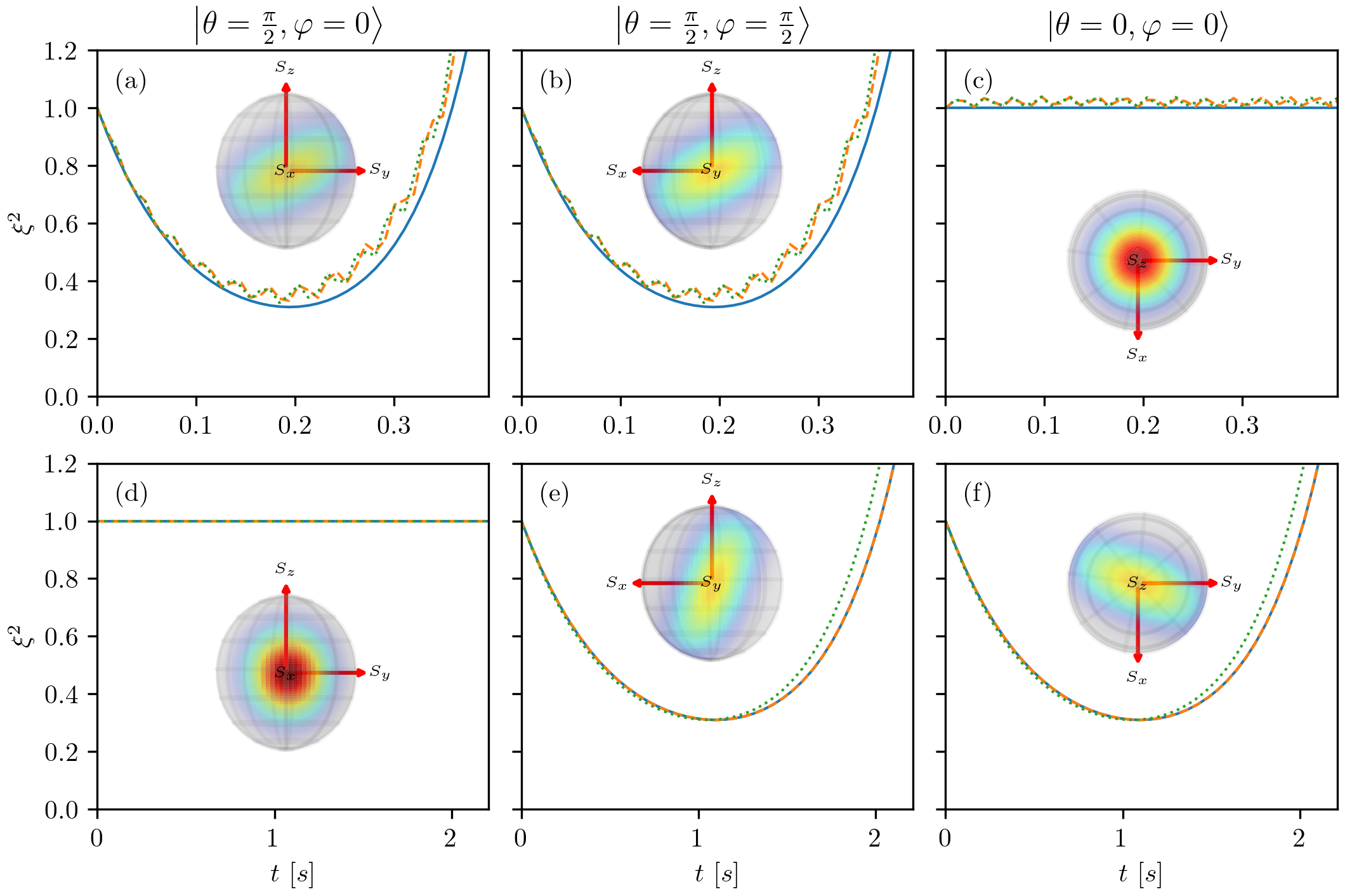}
    \caption{
    Spin squeezing parameter in time (marked by lines) and Husimi distribution projected in the Bloch sphere for $\phi=\pi/5$ (a), (b), (c) and $\phi=\pi$ (d), (e), (f). Spin squeezing generated with FHM (\ref{eq:FHM}) with the coupling term is marked by dotted green lines, spin Hamiltonian by dashed orange lines and effective OAT models by solid blue lines for $N=10$ particles with $U = 1/E_R, J = r U, \Omega = r^3 U$ where $r=1/10$ and $\hbar/E_R = 7 \mu s$.
    The Husimi distribution is calculated using the FHM (\ref{eq:FHM}) at the best squeezing time $t_\text{best}$.
    Panels in each column use the initial state $\ket{\theta, \varphi}$ for angles $\theta,\, \varphi$ as indicated in panels title. Thus, the initial state is oriented along the $x$ direction of the Bloch sphere in the left column;  $y$ direction in the center column and $z$ direction in the right column.
    }
\label{fig:fig2c}
\end{figure}

In this section, we demonstrate by means of exact numerical results how the twisting axis depends on the value of the phase $\phi$.

The time evolution of the spin squeezing parameter for Fermi-Hubbard, spin and effective models using different initial states is shown in Fig.~(\ref{fig:fig2c}). 
The parameters were chosen so that the Fermi-Hubbard model with SFC only approximately simulates the twisting models.
Therefore, the DM term that can appear in Eq.~(\ref{eq:SW-original}) is not completely negligible for $\phi \ne \pi$ corresponding to $\hat{H}^{(2)}_{\rm eff}=\chi_{\phi} \hat{S}_z^2$, as illustrated in panels (a)-(c). On the other hand, for $\phi = \pi$ one has $\hat{H}^{(2)}_{\rm eff}=-\chi_\pi \hat{S}_x^2$, therefore, the dynamics due to the DM term is completely suppressed as it is proportional to $\sin(\phi)$, see panels (d)-(f).
When the initial state corresponds to an eigenstate of the Hamiltonian, the state will only gain a global phase factor and the evolution of spin squeezing parameter will not change. We can see this happening in panel (c) for $\phi=\pi/5$ with initial state $|\theta=0,\varphi = 0\rangle$ which is an eigenstate of $\hat{S}_z$ and in panel (d) for $\phi=\pi$ with initial state $|\theta=\pi/2,\varphi = 0\rangle$ which is an eigenstate of $\hat{S}_x$. 
In contrast, in other panels we can see spin squeezing dynamics because the initial states are not eigenstates of the corresponding effective Hamiltonians. In the case of one-axis twisting Hamiltonian, we obtain the same level of squeezing for any state $|\theta,\varphi \rangle$ if the axis defined by spherical angles $\theta$ and $\varphi$ is perpendicular to the twisting axis, as one can see in the panels (a) and (b) for $\phi \ne \pi$ and  (e) and (f) for $\phi = \pi$. 

A direction of the squeezing, determined by the global sign of the effective Hamiltonian, can be seen from the Husimi distributions $Q(\theta,\varphi)=|\langle \theta,\varphi|\psi(t)\rangle|^2$ for the state $|\psi(t)\rangle = e^{- i t\hat{H}/\hbar} |\theta=\theta_0,\varphi=\varphi_0\rangle$ evolved with the Fermi-Hubbard Hamiltonian (\ref{eq:FHM}) that includes the atom-light coupling term~\cite{1940264}.
Let us focus on panels (b) and (e) in order to compare the tilt of the Husimi distributions.
We remind that panels (b) and (e) illustrate evolution governed by $\hat{H}^{(2)}_{\rm eff} = \chi_{\phi} \hat{S}_z^2$ and $\hat{H}^{(2)}_{\rm eff} = -\chi_\pi \hat{S}_x^2$, respectively.
While the Husimi distribution in panel (b) shows a tilt of the squeezed distribution with respect to the $z$ axis of the Bloch sphere, while panel (e) shows the tilt in the opposite direction with respect to the $x$ axis. This demonstrates the difference in sign of the effective Hamiltonians corresponding to phases $\phi\ne \pi$~and~$\pi$.

\section{Schrieffer--Wolff transformation}

The Schrieffer--Wolff transformation involves a unitary transformation
to perturbatively diagonalize the Hamiltonian of a quantum system
to first order in the interaction. The Schrieffer--Wolff transformation is an operator version of second-order
perturbation theory which is frequently used to project out the high
energy excitations of a specific quantum many-body Hamiltonian $\hat{H}$ to
obtain an effective low energy model.
In below we show a derivation of the general form of the second-order correction that is used by us to derive spin Hamiltonian and effective OAT model.

Consider a quantum system described by the state vector $\left|\Psi \right\rangle $
evolving under the time-dependent Schrödinger equation (TDSE): 
\begin{equation}
\mathrm{i}\hbar\partial_{t}\left|\Psi \right\rangle =\hat{H}\left|\Psi  \right\rangle \,.\label{eq:Schroedinger_Eq}
\end{equation}
The evolution is considered to be governed by the time-independent
Hamiltonian $\hat{H}$ which can be separated into the zero-order Hamiltonian
$\hat{H}_{0}$ and the interaction operator $\hat{V}$, the latter playing a role
of the perturbation: 
\begin{equation}
\hat{H}=\hat{H}_{0}+\hat{V}\label{eq:H-separation}
\end{equation}
It is assumed that the zero order Hamiltonian can be diagonalized
to obtain its eigenstates $\left|m\right\rangle $ and the corresponding
eigen-energies $E_{m}$ 
\[
\hat{H}_{0}\left|m\right\rangle =E_{m}\left|m\right\rangle 
\]
On the other hand, the interaction operator $\hat{V}$ is assumed to be
purely off-diagonal in the basis of eigenstates of $\hat{H}_{0}$, so that
$\left\langle m\right|\hat{V}\left|m\right\rangle =0$ for all $m$. This
assumption does not imply a loss of generality, as any diagonal elements
of $\hat{V}$ can be included to $\hat{H}_{0}$ by modifying its eigenvalues $E_{m}^{\prime}=E_{m}+\left\langle m\right|V\left|m\right\rangle $.
Furthermore, for the reasons that will become clear later, we will
impose a stronger condition assuming that the operator $\hat{V}$ also does
not couple any energy levels with the same energy 
\begin{equation}
V_{nm}=0,\quad\mathrm{for}\quad E_{m}=E_{n}\,.\label{eq:V-condition}
\end{equation}
This condition accommodates the requirement that the operator $\hat{V}$
does not have diagonal matrix elements $V_{nn}=0$.
On the other hand, 
the perturbation operator $\hat{V}$ couples the two energy manifolds spanned by $|m\rangle$ and $|\xi\rangle$, such that
\begin{equation}
    \langle \xi|\hat{V} | m\rangle \ne 0, \,\,\,\,{\rm with}\,\,\, \,\hat{H}_0 |\xi\rangle = E_\xi |\xi\rangle.  
\end{equation}

It is often convenient to go to another representation by applying
a time-independent unitary transformation generated by an anti-Hermitian
operator $\hat{S}$ 
\[
\hat{U}=e^{\hat{S}}\,,\mathrm{with}\quad \hat{S}^{\dagger}=-\hat{S}\,,
\]
so $\hat{U}^{\dagger}\hat{U}=\hat{U}\hat{U}^{\dagger}=1$, as required for the unitary transformation.
Transforming the state vector 
\begin{equation}
\tilde{\left|\Psi\right\rangle }=e^{\hat{S}}\left|\Psi\right\rangle \,,\label{eq:|>-tilde}
\end{equation}
the TDSE takes the form 
\begin{equation}
\mathrm{i}\hbar\partial_{t}\tilde{\left|\Psi \right\rangle }=\tilde{\hat{H}}\tilde{\left|\Psi\right\rangle }\label{eq:Schroedinger_Eq-tilde}
\end{equation}
where 
\begin{equation}
\tilde{\hat{H}}=e^{\hat{S}}\hat{H}e^{-\hat{S}}\,\label{eq:H-tilde}
\end{equation}
is the transformed Hamiltonian. It can be expanded in an infinite
series in the powers of the operator $\hat{S}$ in a usual way:
\begin{equation}
\tilde{\hat{H}}=\hat{H}+\left[\hat{S},\hat{H}\right]+\frac{1}{2}\left[\hat{S},\left[\hat{S},\hat{H}\right]\right]+\ldots\,,\label{eq:H-tilde-expanded}
\end{equation}
where $\left[\hat{A},\hat{B}\right]=\hat{A}\hat{B}-\hat{B}\hat{A}$ labels a commutator between two operators
$A$ and $B$. Generally one should sum the whole series. Yet, if
the generator of the unitary transformation $\hat{S}$ is small enough,
one can truncate the expansion at a finite order of expansion series,
often at the first or the second order.

In the Schrieffer--Wolff transformation the exponent $\hat{S}$ of the
unitary transformation is chosen such that the transformed Hamiltonian
$\tilde{\hat{H}}$ given by Eq.(\ref{eq:H-tilde-expanded}) is diagonal
in the eigenstates of $\hat{H}_{0}$ up to second order in the perturbation
$V$, rather than up to the first order of perturbation, as in the
original Hamiltonian (\ref{eq:H-separation}). For this purpose, let
us rewrite the expansion (\ref{eq:H-tilde-expanded}) of the transformed
Hamiltonian in terms of of the zero-order Hamiltonian $\hat{H}_{0}$ and
the perturbation $V$: 
\begin{equation}
\tilde{\hat{H}}=\hat{H}_{0}+\hat{V}+\left[\hat{S},\hat{H}_{0}\right]+\left[\hat{S},\hat{V}\right]+\frac{1}{2}\left[\hat{S},\left[\hat{S},\hat{H}_{0}\right]\right]+\frac{1}{2}\left[\hat{S},\left[\hat{S},\hat{V}\right]\right]+\ldots\label{eq:H-tilde-expanded-H_0,V}
\end{equation}
The Hamiltonian $\tilde{\hat{H}}$ can be made diagonal to first order in
$\hat{V}$ (i.e. up to the second order in $\hat{V}$) by choosing the generator
$\hat{S}$ such that 
\begin{equation}
\left[\hat{H}_{0},\hat{S}\right]=\hat{V}\,.\label{eq:Commutator_H_0,S-zero}
\end{equation}
Substituting this choice to the expansion (\ref{eq:H-tilde-expanded-H_0,V}),
in which $\left[\hat{S},\left[\hat{S},\hat{H}_{0}\right]\right]=-\left[\hat{S},\hat{V}\right]$,
one arrives at: 
\begin{equation}
\tilde{\hat{H}}=\hat{H}_{0}+\frac{1}{2}\left[\hat{S},\hat{V}\right]+\frac{1}{2}\left[\hat{S},\left[\hat{S},\hat{V}\right]\right]+\ldots\label{eq:H-tilde-SW}
\end{equation}
Equation (\ref{eq:Commutator_H_0,S-zero}) represents a condition
for the Schrieffer--Wolff transformation providing the transformed
Hamiltonian given by Eq.(\ref{eq:H-tilde-SW}). The condition (\ref{eq:Commutator_H_0,S-zero})
can be rewritten in terms of the matrix elements
\begin{equation}
\left(E_{\xi}-E_{m}\right)S_{\xi m}=V_{\xi m}\,.\label{eq:Commutator_H_0,S-zero-1}
\end{equation}
This defines matrix elements of the generator $S_{\xi m}$ as long as
$E_{m}\ne E_{\xi}$. That's why we have imposed the condition (\ref{eq:V-condition})
assuming that the operator $V$ does not couple degenerate energy
levels. Using the assumption (\ref{eq:V-condition}), the condition
(\ref{eq:Commutator_H_0,S-zero}) is fulfilled by taking the following
matrix elements of the generator $\hat{S}$ 
\begin{equation}
S_{\xi m}=\left\{ \begin{array}{c}
\frac{V_{\xi m}}{E_{\xi}-E_{m}}\,,\quad\mathrm{for}\quad E_{m}\ne E_{\xi}\\
0\,,\quad\quad\quad\,\mathrm{for}\quad E_{m}=E_{\xi}
\end{array}\,.\right.\label{eq:S_nm-explicit}
\end{equation}
In this way, the matrix elements are $S_{\xi m}\propto V_{\xi m}$, so expansion
in the powers of the generator $\hat{S}$ implies its expansion in the powers
of the perturbation $\hat{V}$. Substituting Eq.~(\ref{eq:S_nm-explicit})
into (\ref{eq:H-tilde-SW}), the matrix elements of the transformed
Hamiltonian read up to the second order term in the perturbation $V$
\begin{equation}
\tilde{H}_{n m}\approx E_{n}\delta_{nm}+\frac{1}{2}\sum_{\xi}\left(\frac{V_{n \xi}V_{\xi m}}{E_{n}-E_{\xi}}+\frac{V_{n \xi}V_{ \xi m}}{E_{m}-E_{\xi}}\right)\,,\quad\mathrm{with}\quad E_{\xi}\ne E_{n},E_{m}\,,\label{eq:H-tilde-SW-exclicit-matr.el.}
\end{equation}
where the condition $E_{\xi}\ne E_{n},E_{m}$ excludes from the summation
the terms with $E_{\xi}=E_{n}$. The diagonal elements 
\[
\tilde{H}_{nn}=E_{n}+\sum_{E_{\xi}\ne E_{n}}V_{n \xi}V_{\xi n}/\left(E_{n}-E_{\xi}\right)
\]
represent the eigen-energies containing the usual second order corrections.
On the other hand, the off-diagonal elements $\tilde{H}_{nm}$ describe
coupling between the different energy manifolds.

The matrix representation of the second-order correction~(\ref{eq:H-tilde-SW-exclicit-matr.el.}) was used by us in Section~\ref{app:SWT} of Supplementary Materials to calculate the 
projection of the Fermi-Hubbard model with atom-light coupling over the single occupancy manifold. 

Note that the second order correction~(\ref{eq:H-tilde-SW-exclicit-matr.el.}) can be written using projection operators as
\begin{equation}
    \tilde{\hat{H}} \approx \hat{I} \hat{H}_0 \hat{I} + \hat{I} \hat{V} \hat{G} \hat{V} \hat{I},
\end{equation}
where $\hat{I}=\sum_m |m\rangle \langle m|$ is a projector operator over the energy manifold spanned by $|m\rangle$, and $\hat{G}=\sum_\xi \frac{|\xi\rangle \langle \xi|}{\Delta E}$ is an operator which sums projections over the energy manifold spanned by $|\xi\rangle$ with the corresponding energy mismatch denominator $\Delta E$. The operator level of second-order correction was used by us in the main text to derive the effective OAT model.

\end{widetext}

\bibliography{bibliography}

\end{document}